\documentclass[12pt]{amsart}

\usepackage{amsmath,amssymb,dsfont}
\usepackage{graphicx}

\newfont{\bbd}{msbm10 scaled\magstep1}

\def\id{\hbox{{1}\kern-.25em\hbox{\rm l}}}
\def\one#1{#1^{\raise5pt\hbox{$\scriptstyle\!\!\!\!1$}}\,{}}
\def\two#1{#1^{\raise5pt\hbox{$\scriptstyle\!\!\!\!2$}}\,{}}

\def\comment#1{}

\def\?{(?)\marginpar{|?}}

\def\beq{\begin{equation}}
\def\eeq{\end{equation}}
\def\bea{\begin{eqnarray}}
\def\eea{\end{eqnarray}}
\def\bmat{\left(\begin{array}}
\def\emat{\end{array}\right)}

\newcounter{subequation}[equation]
\makeatletter
\expandafter\let\expandafter
\reset@font\csname reset@font\endcsname
\def\subeqnarray{\arraycolsep1pt
    \def\@eqnnum\stepcounter##1{\stepcounter{subequation}%
        {\reset@font\rm(\theequation\alph{subequation})}}
\jot5mm     \eqnarray}

\makeatother

\newcounter{appendix}
\def\mat2#1#2#3#4{{\left(\begin{array}{cc}#1 & #2\\ #3 & #4
      \end{array}\right)}}

\def\mats2#1#2#3#4{{\left(\begin{array}{cc}#1 & #2\vspace{2truemm} \\ #3 & #4
\end{array}\right)}}

\begin{document}

\title{Multiscale expansion of the lattice Sine--Gordon equation}
\author{Xiaoda Ji}

\address{Department of Mathematics\\ University of Science and Technology of
China, Hefei, PeopleÕs Republic of China\\
E-mail: jxd@ustc.edu.cn}

\author{D. LEVI}

\address{Dipartimento di Ingegneria Elettronica \\
Universit\`a degli Studi Roma Tre and Sezione INFN, Roma Tre \\
Via della Vasca Navale 84,00142 Roma, ITALY\\
E-mail: levi@fis.uniroma3.it}

\author{M. PETRERA}

\address{Dipartimento di Fisica E. Amaldi \\
Universit\`a degli Studi Roma Tre and Sezione INFN, Roma Tre \\
Via della Vasca Navale 84,00142 Roma, ITALY\\ 
E-mail: petrera@fis.uniroma3.it}

\begin{abstract}
We expand a discrete--time lattice sine--Gordon equation on multiple
lattices and obtain the partial difference equation  which governs its far field behaviour.
 Such reduction allow us to obtain a new completely discrete
nonlinear Schr\"oedinger (NLS) type equation.
\end{abstract}

\maketitle

\section{Introduction}
Reductive perturbation technique \cite{t1,t2} has proved to be an important tool to find approximate solutions for many important physical problems by reducing the given nonlinear partial differential equation in the far field often to an integrable one. Recently this approach has been extended to the case of equations living on lattices \cite{levi,lp}. Here we apply it to the case of a discrete--time lattice sine--Gordon equation.

In Section \ref{sm} we briefly describe the discrete  perturbation technique and in Section \ref{mkdv} we apply it to a discrete--time lattice sine--Gordon equation.

\section{The discrete  perturbation technique}  \label{sm}

The aim of this section is to fix the notation and to introduce the
formulae necessary to reduce lattice equations in the
framework of the discrete reductive perturbation technique \cite{levi,lp}.

Given a lattice defined by a constant spacing $h$, we will
 denote by $n$ the running index of the points separated by $h$.
In correspondence with the lattice variable $n$,
we can introduce the real variables $x = h \,n$.

We can define  on the same lattice a set of slow varying variables  by  introducing a large integer number $N$, defining a small parameter $\epsilon=N^{-1}$ and requiring that
\begin{eqnarray} \label{f2b}
n_j = \epsilon^j \, n.
\end{eqnarray}
This correspond to sampling points from the original lattice which are situated at a distance of $N^j  \,h$ between them. If we set them on a lattice of spacing $h$, the corresponding slowly varying real variables $x_j$ are related to the variable $x$ by the equation $x_j = \epsilon^j \, x$.

Let us
 consider a function $f\doteq f_n$ defined on the points of a lattice variable $n$ and
let us assume
that $f_n = g_{n_1,n_2, \dots,n_K}$, i.e. $f$ depends on a finite
number $K$ of slow varying lattice variables $n_j$ $j=1,2,\ldots,K$ defined as in  eq. (\ref{f2b}).
 We are looking for explicit expressions for, say,  $f_{n+1}$ in terms of
$g_{n_1, n_2, \ldots, n_K}$ evaluated
on the points of the $n_1$, $n_2$, $\ldots, n_K$ lattices.
At first let us consider the case, studied by Jordan \cite{Jordan}, when we have only two different lattices,
i.e. $K = 1$. Using the results obtained in this case we will then
consider the case corresponding to $K = 2$. The general case will than be obvious.

\vspace{.5cm}

{\bf{I)}} $K=1$ ($f_n=g_{n_1}$).
In Jordan \cite{Jordan} we find the following formula:
\begin{equation} \label{difInt}
\Delta^k \, g_{n_1}\doteq \sum_{i=0}^{k}  (-1)^{k-i}
{k \choose i} g_{n_1+i} =\sum_{i=k}^\infty  \frac{k!}{i!}P(i,k)\, \Delta^i  f_n.
\end{equation}
Here  the coefficients
$P(i,k)$ are given by
\begin{eqnarray} \label{p1}
P(i,k)=\sum_{\alpha=k}^i
 \omega^\alpha S_{i}^{\alpha} \, \mathfrak{S}_{\alpha}^k,
\end{eqnarray}
where $\omega$ is the ratio of the increment in the lattice  variable $n$
with respect to that of variable $n_1$. In this case, taking into account eq. (\ref{f2b}),
$\omega=N$.
The coefficients $S_{i}^{\alpha}$ and $\mathfrak{S}_{\alpha}^k$ are the Stirling numbers of the
first and second kind respectively \cite{AS}. Formula (\ref{difInt}) allow us to express a difference of
order $k$ in the lattice variable $n_1$ in terms of an infinite
number of differences on the lattice variable $n$. The result (\ref{difInt}) can be inverted and we get:
\begin{equation} \label{difInt1}
\Delta^k \, f_{n}=\sum_{i=k}^\infty  \frac{k!}{i!}Q(i,k)\, \Delta^i  g_{n_1},
\end{equation}
where  the coefficients
$Q(i,k)$ are given by (\ref{p1}) with $\omega=N^{-1}=\epsilon$.

To get from eq. (\ref{difInt})
a finite approximation of the variation of $g_{n_1}\doteq f_n$  we need
to truncate the expansion in the r.h.s.
by requiring a slow varying condition for the function $f_n$.
Let us introduce the following definition:

\vspace{.2cm}

{\bf{Definition}}. $f_n$ is a slow varying function of
order $p$ iff $\Delta^{p+1} \, f_n=0$.

\vspace{.2cm}

From the above definition  it follows that a slow varying function of
order $p$ is a polynomial of degree $p$ in $n$. From eq.
(\ref{difInt}) we see that if $f_n$ is a slow varying function of
order $p$ then $\Delta^{p+1} \, g_{n_1}=0$, namely $g_{n_1}$ is also
 of order $p$.
Eq. (\ref{difInt1}) provide us with  the formulae for
$f_{n+1}$ in terms of $g_{n_1}$  and its neighboring points in the case of slow varying functions of any order $p$. Let us write down explicitly these expressions in the case of $g_{n_1}$ of order 1 and 2.

\begin{itemize}
\item $p=1$. Formula (\ref{difInt1}) reduces to
\begin{equation} \nonumber
\Delta f_{n} =\frac{1}{N}\Delta  g_{n_1},
\end{equation}
i.e. $f_{n+1}$ reads
\begin{equation} \nonumber
f_{n+1} = g_{n_1} + \frac{1}{N} (g_{n_1+1}- g_{n_1})+ O(N^{-2}).
\end{equation}

\item $p=2$. From eq. (\ref{difInt1}) we get
\begin{equation} \nonumber
\Delta f_{n} = \frac{1}{N} \Delta  g_{n_1} + \frac{1-N}{2\,  N^2}
\Delta^2 \, g_{n_1},
\end{equation}
and thus $f_{n+1}$ reads
\begin{eqnarray}
f_{n+1} &=& g_{n_1} + \frac{1}{2\,N} (-g_{n_1+2}+4 \,g_{n_1+1}-3 \,g_{n_1})+ \nonumber \\
 && + \frac{1}{2\,N^2} (g_{n_1+2}-2 g_{n_1+1}+g_{n_1})+
O(N^{-3}). \label{p=22}
\end{eqnarray}
\end{itemize}

In the next section we will consider the reduction of an integrable lattice sine--Gordon
equation. It is known \cite{ly} that a scalar
differential--difference equation can possess higher conservation laws and thus be integrable only if
it depends symmetrically on the discrete variable.
The results contained in (\ref{difInt1}) do not provide us with symmetric
formulae. To get symmetric formulae we  take into account the
following remarks:
\begin{enumerate}
\item \label{r1} Formula (\ref{difInt}) holds also if  $h$ is negative;
\item \label{r2} For a slow varying function of order $p$, we have $\Delta^p  f_n = \Delta^p  f_{n +\ell}$,
for all $\ell \in \mathbb{Z}$.
\end{enumerate}

When  $f_n$ is a slow varying function of odd order we are not able
to construct completely symmetric derivatives using
just an odd number of points centered around the $n_1$ point and thus $f_{n \pm 1}$ can never be expressed
in a symmetric form.

From eq. (\ref{difInt1}), using the above remarks we can easily construct the symmetric version of (\ref{p=22}).
We get:
\begin{equation} \label{p23}
f_{n + 1} =
g_{n_1}+  \frac{1}{2\, N}(g_{n_1+1} - g_{n_1-1}) +
\frac{1}{2\, N^2}(g_{n_1+1}- 2 \,g_{n_1}+ g_{n_1-1})+
O(N^{-3}).
\end{equation}

\vspace{.5cm}

{\bf{II)}} $K=2$ ($f_n=g_{n_1,n_2}$). The derivation of the formulae in this case is done in
the same spirit as for the symmetric expansion presented above, resulting in eq.
(\ref{p23}). Let us just consider the case when $p=2$,
as this is the lowest value of $p$ for which we can consider
$f_n$ as a function of the two scales $n_1$ and $n_2$.  From eq.
(\ref{difInt1}) we get:
\begin{eqnarray} \label{p24}
&& g_{n_1+1,n_2} = g_{n_1,n_2} + N \Delta_1 f_{n,n} + \frac{1}{2} N ( N-1) \Delta_1^2 f_{n,n}, \\ \label{p25}
&& g_{n_1,n_2+1} = g_{n_1,n_2} + N^2 \Delta_2 f_{n,n} + \frac{1}{2} N^2 ( N^2-1) \Delta_2^2 f_{n,n}.
\end{eqnarray}
Here the symbols $\Delta_1$ and $\Delta_2$
denote difference operators which acts on the first and respectively on the second index
of the function $f_{n,n} \doteq g_{n_1,n_2}$, e.g.  $\Delta_1 f_{n,n} \doteq f_{n+1,n} - f_{n,n}$.
and $\Delta_2 f_{n,n} \doteq f_{n,n+1} - f_{n,n}$.

Let us now consider a function $g_{n_1,n_2}$
where one shifts both indices by $1$. From eq. (\ref{p24}), taking into account that, from eq. (\ref{f2b}), for example, $g_{n_1+1,n_2} = f_{n+N,n}$,  one has:
\begin{eqnarray} \label{p26}
g_{n_1+1,n_2+1} = g_{n_1,n_2+1} + N \Delta_1 f_{n,n+N^2} + \frac{1}{2} N ( N-1) \Delta_1^2 f_{n,n+N^2}.
\end{eqnarray}
Using the result (\ref{p25}) we can write eq. (\ref{p26}) as
\begin{eqnarray}
g_{n_1+1,n_2+1}
&=&  g_{n_1,n_2} + N^2 \Delta_2 f_{n,n} + \frac{1}{2} N^2 ( N^2-1) \Delta_2^2 f_{n,n} + \nonumber  \\
&& + N  \Delta_1  f_{n,n} + N^3 \Delta_1 \Delta_2 f_{n,n} +
\frac{1}{2} N^3 ( N^2-1) \Delta_1 \Delta_2^2 f_{n,n} + \nonumber \\
&& + \frac{1}{2} N ( N-1) \Delta_1^2  f_{n,n} +
N^3 ( N-1) \Delta_1^2 \Delta_2 f_{n,n} + \nonumber \\
&& +\frac{1}{4} N^3 ( N^2-1)( N-1) \Delta_1^2 \Delta_2^2 f_{n,n}. \label{p27}
\end{eqnarray}
As, using the second remark, the second difference of $f_{n,n}$ depends just on
its nearest neighboring points, the right hand side of eq. (\ref{p27}) depends,
apart from $f_{n,n}=g_{n_1,n_2}$, on
$f_{n,n+1}, \, f_{n,n-1}, \, f_{n+1,n}, \, f_{n-1,n}$, $ f_{n+1,n+1}, \, f_{n+1,n-1}, \, f_{n-1,n+1}, $
and $f_{n-1,n-1}$, i.e. 8 unknowns. Starting from eqs. (\ref{p24}), (\ref{p25}) and (\ref{p27})
we can write down 8 equations, using the first remark,
which define $g_{n_1+1,n_2}$, $g_{n_1-1,n_2}$, $g_{n_1,n_2+1}$, $g_{n_1,n_2-1}, \, g_{n_1+1,n_2+1}$,
$g_{n_1+1,n_2-1}, \, g_{n_1-1,n_2+1},$ and $g_{n_1-1,n_2-1}$ in terms of  the
functions $f_{n+i,n+j}$ with $(i,j)=0,\pm1$. Inverting this system of
equations we get $f_{n \pm 1}$ in term of $g_{n_1,n_2}$ and its shifted values:
\begin{eqnarray}
f_{n \pm 1} &=& g_{n_1,n_2} \pm  \frac{1}{ 2\, N} (g_{n_1+1,n_2}   - g_{n_1-1,n_2})
\pm  \frac{1}{2 \,N^2}(g_{n_1,n_2+1} - g_{n_1,n_2-1})+ \nonumber \\
&& + \frac{1}{2 \,N^2} (g_{n_1+1,n_2} - 2 \,g_{n_1,n_2} + g_{n_1-1,n_2}) + \nonumber \\
&& +  \frac{1}{4\, N^3}
( g_{n_1+1,n_2+1} - g_{n_1-1,n_2+1} - g_{n_1+1,n_2-1} + g_{n_1-1,n_2-1}) + \nonumber \\
&& + O(N^{-4}).  \label{p27b}
\end{eqnarray}
It is worthwhile to notice that the two lowest order (in $N^{-1}$) terms of the expansion (\ref{p27b})
are just the sum of the first symmetric differences of $g_{n_1}$ and $g_{n_2}$.
Thus in the continuous limit, when we divide by $h$ and send $h$ to zero in
such a way that $x=h\,n$, $x_1=h\,n_1$ and $x_2=h\,n_2$ be finite, we will have
$
f_{,x} = \epsilon \,g_{x_1} + \epsilon^2 \,g_{x_2}
$.
Extra terms appear at the order $N^{-3}$ and contain shifts in both $n_1$ and $n_2$.

When $f_n$ is a slow varying function of order 2 in $n_1$  it can also be
 of order 1  in $n_2$. In such a case eq. (\ref{p25}) is given by
\begin{eqnarray} \label{p28}
g_{n_1,n_2+1} = g_{n_1,n_2} + N \Delta_2 f_{n,n}.
\end{eqnarray}
Starting from eqs. (\ref{p24}), (\ref{p28})  and a modified (\ref{p27}) we can get a set of 8 equations which allows us
to get $f_{n \pm 1}$ in terms of $g_{n_1,n_2}$ and its shifted values.  In such a case
 $f_{n \pm 1}$ reads
\begin{eqnarray}
f_{n \pm 1} &=& g_{n_1,n_2} \pm \frac{1}{ 2 \,N} (g_{n_1+1,n_2}  - g_{n_1-1,n_2})
+ \frac{1}{ N^2} (g_{n_1,n_2 \pm 1} - g_{n_1,n_2} )  + \nonumber \\
&& + \frac{1}{2 \,N^2}(g_{n_1+1,n_2} - 2 \,g_{n_1,n_2} + g_{n_1-1,n_2})+
O(N^{-3}). \label{h16}
\end{eqnarray}

It is possible to introduce two parameters in the definition of $n_1$, $n_2$ in terms of $n$. Let us
define
$
n_1 \doteq (n \,M_1)/N, n_2 \doteq (n \,M_2)/N^2
$,
where $M_1$ and $M_2$ are divisors of $N$ and $N^2$ so that $n_1$ and $n_2$  are integers numbers.
In such a case eq. (\ref{p27b}) reads
\begin{eqnarray}
f_{n \pm 1} &=& g_{n_1,n_2} \pm  \frac{M_1}{ 2 N} (g_{n_1+1,n_2}   - g_{n_1-1,n_2})
\pm  \frac{M_2}{2 \,N^2}(g_{n_1,n_2+1} - g_{n_1,n_2-1})+ \nonumber  \\
&&+ \frac{M_1^2}{2 \,N^2} (g_{n_1+1,n_2} - 2 \,g_{n_1,n_2} + g_{n_1-1,n_2})
+ \nonumber  \\
&& +  \frac{M_1 M_2}{4 \,N^3}
( g_{n_1+1,n_2+1} - g_{n_1-1,n_2+1} - g_{n_1+1,n_2-1} + g_{n_1-1,n_2-1}) + \nonumber  \\
&& + O(N^{-4})  \nonumber
\end{eqnarray}
and eq. (\ref{h16}) accordingly.

 Let us consider  the case of  two independent lattices and a function $f_{n,m}$ defined on them. As the two lattices are independent the formulae
presented above apply independently
on each of the lattice variables. So, for instance, the variation $f_{n+1,m}$
when the function $f_{n,m}$ is a slowly varying function of order 2 of a lattice variable $n_1$ reads
\begin{eqnarray}
f_{n+1,m} &=&g_{n_1,m} +  \frac{1}{2 \,N} (g_{n_1+1,m} - g_{n_1-1,m}) + \nonumber  \\
&& + \frac{1}{2 \,N^2}(g_{n_1+1,m} - 2 \,g_{n_1,m}+ g_{n_1-1,m}) +
O(N^{-3}).\nonumber
\end{eqnarray}

A slightly less obvious situation appears when we consider $f_{n+1,m+1}$, as new terms will appear, see Levi et al. \cite{levi,lp} for the formulae in this case.

\section{Reduction of the lattice sine--Gordon equation} \label{mkdv}

A discrete analogue of the sine--Gordon equation is given by the
following nonlinear P$\Delta$E \cite{frank1}:
\begin{equation} \label{3.1}
u_{n+1,m+1}=\frac{1}{u_{n,m}}\, \frac{u_{n+1,m} \, u_{n,m+1} -p^4}{1- q^4 \, u_{n+1,m} \, u_{n,m+1}}.
\end{equation}
This equation involves just four points which lay on two
orthogonal infinite lattices and are the vertices of an elementary square. When written in polynomial form has quartic nonlinearity.
In eq. (\ref{3.1})  $u_{n,m}$ is the dynamical (real) field variable at site 
$(m,n) \in \mathbb{Z} \times \mathbb{Z}$ and
 $p,q \in \mathbb{R}$ are the lattice parameters. These are
assumed different from zero and will go to zero in the continuous limit
so as to get the continuous sine--Gordon equation.

To get a nonlinear dispersion relation we carry out the change of variable
$
u_{n,m} \mapsto
p/q + u_{n,m}.
$ 
The linear part of the resulting equation is given by:
\begin{equation} \label{3.2a}
(\sigma -1)(u_{n,m}+ u_{n+1,m+1})+(\sigma +1)(u_{n+1,m}+u_{n,m+1})=0,
\end{equation}
where $\sigma \doteq p^2 q^2$.

The general solution of eq. (\ref{3.2a}) is written as a superposition of linear waves
$
E_{n,m} = {\rm{exp}}[{\rm {i}}(\,k\, n - \, \omega(k)\,m)] \doteq  z^n \, \Omega^m.
$
The dispersion relation for these linear waves is given by
\begin{equation} \label{dr}
\Omega = e^{-{\rm {i}} \omega } =-\frac{(\sigma+1)\, z +\sigma-1}{(\sigma-1)\,z +\sigma+1},
\end{equation} 
namely
$$
\omega = -\arctan \left[ \frac{2\, \sigma \sin k}{(\sigma^2+1)\cos k +\sigma^2-1}\right].
$$
From eq. (\ref{dr}),
by differentiation with respect to $k$, we get the group velocity:
\begin{equation} \label{l1}
\omega_{,k} = - \frac{4 \, \sigma \, z}{[(\sigma+1)\,z +\sigma-1] [(\sigma-1)\,z +\sigma+1]}.
\end{equation}

We now look for real solutions of the nonlinear equation (\ref{3.1}) written as a combination of modulated waves:
\begin{equation} \label{3.9}
u_{n,m} = \sum_{s=0}^{\infty} \epsilon^{\beta_s}
\psi^{(s)}_{n,m}\,  (E_{n,m})^s +
\sum_{s=1}^{\infty} \epsilon^{\beta_s}
{\bar \psi}^{(s)}_{n,m} \, ({\bar E}_{n,m})^s,
\end{equation}
where the functions $\psi^{(s)}_{n,m}$ are  slowly varying functions on the lattice,
i.e. $\psi^{(s)}_{n,m}=\psi^{(s)}_{n_1,m_1,m_2}$  and $\epsilon^{\gamma} = N^{-1}$.
By  $\bar b$ we mean the complex conjugate of a complex quantity $b$ so that, for example,
${\bar E}_{n,m} = (E_{n,m})^{-1}$.
The positive numbers $\beta_s$ are to be determined in such a way that :
\begin{enumerate}
\item $\beta_1 \leq \beta_s \; \forall \,  s=0,2,3,\dots,\infty$.
In general it is possible to set $\beta_1=1$.
\item In the equation for $\psi^{(1)}_{n,m}=\psi_{n,m}$,  the lowest order  nonlinear terms should match  the slow
time derivative of the linear part after having solved all linear equations. This will provide a relation between $\gamma$ and the $\beta_s$.
\end{enumerate}

Introducing the expansion (\ref{3.9})
in the P$\Delta$E obtained from equation (\ref{3.1}) after the change of variable
$
u_{n,m} \mapsto
p/q + u_{n,m}
$,
we analize the coefficients of the various harmonics $(E_{n,m})^s$
for $s=1$, $s=2$ and $s=0$ and, as assuming that $\beta_s$ increases with $s$,  the nonlinear terms
will depend only on the lowest $s$ terms, we came to the conclusion that we can choose
$
\gamma=1, \beta_0 = 2, \beta_s = s, s \ge 1.
$
The discrete slow varying variables $n_1$, $m_1$ and $m_2$ are defined in terms of $n$ and $m$ by:
\begin{equation} \nonumber
n_1 \doteq  \frac{M_1\, n}{N}, \qquad m_1 \doteq  \frac{M_2\, m}{N}, \qquad m_2  \doteq \frac{n}{N^2}.
\end{equation}

For $s=1$
we get, at the lowest order in $\epsilon$, a linear equation
which is identically solved by the dispersion relation (\ref{dr}).

At $\epsilon^2$ we get the linear equation
\begin{eqnarray}
&& M_1 \,z \,  [ (\sigma-1)\, \Omega+\sigma+1]\,
(\psi_{n_1+1,m_1,m_2} - \psi_{n_1-1,m_1,m_2} )+  \nonumber \\
&+&M_2 \,\Omega \, [ (\sigma-1)\, z+\sigma+1] \, (  \psi_{n_1,m_1+1,m_2} - \psi_{n_1,m_1-1,m_2} )
= 0, \label{3.11du}
\end{eqnarray}
whose solution is given by
$
\psi_{n_1,m_1,m_2} = \phi_{n_2,m_2}, \,  n_2 = n_1 - m_1
$,
provided that the integers $M_1$ and $M_2$ are choosen as
\begin{eqnarray} \label{m1m2}
M_1 = S \, \Omega \, [ (\sigma-1)\,z+\sigma+1], \quad
M_2 = S \, z \, [ (\sigma-1)\,\Omega+\sigma+1],
\end{eqnarray}
where $S \in \mathbb{C}$ is a constant. Defining
$S = \rho \exp( {{\rm{i}}\, \theta})$, $\rho \in \mathbb{R}_+$ and
$-\pi \leq \theta < \pi$,
we can choose $\theta$ and $\rho$ in such a way that $M_1$ is an integer number:
\begin{equation} \nonumber
\theta = -\arctan \left[
\frac{(\sigma+1) \sin k}{(\sigma+1)\cos k +\sigma-1}
\right]+ \ell\, \pi, \qquad \ell \in \mathbb{Z},
\end{equation}
\begin{eqnarray}
\rho= (-1)^\ell M_1 \, \frac{1}{\sqrt{2}}\, [(\sigma^2-1) \cos k +\sigma^2 +1]^{-1/2}. \nonumber
\end{eqnarray}
The request that also $M_2$ is an integer impose a constraint on $k$ as $M_2 = \omega_{,k} M_1$,
i.e.  $\omega_{,k} \in \mathbb{Q}$, see eqs. (\ref{l1}) and (\ref{m1m2}).
Let us notice that also $n_2 = n_1 + m_1$ solves eq. (\ref{3.11du}) by an appropriate choice of
$M_1$ and $M_2$.

At $\epsilon^3$ we get a nonlinear equation for $\phi_{n_2,m_2}$ which
depends on $\psi^{(0)}_{n_2,m_2}$ and $\psi^{(2)}_{n_2,m_2}$:
\begin{eqnarray} \label{yyk}
&& \phi_{n_2,m_2+1} - \phi_{n_2,m_2} +
   c_1 \, ( \phi_{n_2+2,m_2} + \phi_{n_2-2,m_2} - 2 \, \phi_{n_2,m_2} )  + \nonumber  \\
&+& c_2 \,( \phi_{n_2+1,m_2} +
\phi_{n_2-1,m_2} - 2 \, \phi_{n_2,m_2} ) +
c_3  \, \phi_{n_2,m_2} |{\phi}_{n_2,m_2}|^2+  \nonumber  \\
&+& c_4  \, \psi^{(0)}_{n_2,m_2} \, {\phi}_{n_2,m_2}+
c_5  \, \psi^{(2)}_{n_2,m_2} \, \bar{{\phi}}_{n_2,m_2} = 0,
\end{eqnarray}
where the $c_i$'s, $1\leq i\leq 5$,
are known coefficients depending on
$z$, $S$ and the lattice parameters $p,q$.

The functions $\psi^{(0)}_{n_2,m_2}$ and $\psi^{(2)}_{n_2,m_2}$ that appear in equation
(\ref{yyk}) are obtained considering the equations for the harmonics
$s=0$, at the third order in $\epsilon$, and $s=2$ at the second one. We get:
\begin{equation}
\psi^{(0)}_{n_2,m_2}=\frac{q}{p}\,|{\phi}_{n_2,m_2}|^2, \quad
\psi^{(2)}_{n_2,m_2}=\frac{1}{2} \,  \frac{q}{p}\,({\phi}_{n_2,m_2})^2. \label{sot}
\end{equation}
Inserting $\psi^{(0)}_{n_2,m_2}$ and $\psi^{(2)}_{n_2,m_2}$ given by
eqs. (\ref{sot}) in equation (\ref{yyk}) we obtain the following nonlinear
lattice equation:
\begin{eqnarray}
&& {\rm{i}} \, (\phi_{n_2,m_2+1} - \phi_{n_2,m_2}) +
   \hat c_1 \, ( \phi_{n_2+2,m_2} + \phi_{n_2-2,m_2} - 2 \, \phi_{n_2,m_2} )  + \nonumber  \\
&+& \hat c_2 \,( \phi_{n_2+1,m_2} +
\phi_{n_2-1,m_2} - 2 \, \phi_{n_2,m_2} ) +
\hat c_3  \, \phi_{n_2,m_2} |{\phi}_{n_2,m_2}|^2= 0, \label{yyy}
\end{eqnarray}
where the coefficients $\hat c_i$'s, $1\leq i\leq 3$,  can be computed also recalling that
$z \doteq {\rm{exp}}({\rm{i}} \, k) $. They read:
\begin{eqnarray}
\hat c_1 &=&
{\rm {i}} \,
\frac{M_2^2 \,(\sigma-1)[(\sigma+1)(\cos k +{\rm{i}} \sin k) +\sigma+1]}{16 \, \sigma}, \nonumber \\
\hat c_2 &=& - {\rm {i}} \, \frac{M_2^2 \,(\sigma-1)[(\sigma+1) \cos k +\sigma+1]}{4 \, \sigma},
\label{ccc} \\
\hat c_3 &=& \frac{2  \,q^4  \,(\sigma^2-1)  \sin^3 k }{[(\sigma^2-1)  \cos k +\sigma^2+1)]}.
\nonumber
\end{eqnarray}
The coefficients (\ref{ccc}) depend just on  the integer constant $M_2$.

Taking into account that $\phi_{n_2,m_2}$ is a slow varying function of order $2$ in $n_2$ we can, using the remark \ref{r2}, substitute $\phi_{n_2+2,m_2}$ by $3 \,\phi_{n_2+1,m_2} -3 \,  \phi_{n_2,m_2} + \phi_{n_2-1,m_2}$ and $\phi_{n_2-2,m_2}$ by $3\,  \phi_{n_2-1,m_2} -3 \, \phi_{n_2,m_2} + \phi_{n_2+1,m_2}$. In such a way eq. (\ref{yyy}) becomes
 \begin{eqnarray}
 {\rm{i}} \, (\phi_{n_2,m_2+1} - \phi_{n_2,m_2}) &+&
  (4 \hat c_1 + \hat c_2 )\,( \phi_{n_2+1,m_2} +
\phi_{n_2-1,m_2} - 2 \, \phi_{n_2,m_2} ) + \nonumber \\ &+&
\hat c_3  \, \phi_{n_2,m_2} |{\phi}_{n_2,m_2}|^2= 0, \label{yyyf}
\end{eqnarray}
where
\begin{eqnarray} \label{yyyy2}
4\,\hat c_1 +\hat c_2 = -\frac{M^2_2\, (\sigma^2-1 ) \sin k}{4  \, \sigma}.
\end{eqnarray}

The P$\Delta$E (\ref{yyyf}) is a {\it completely discrete  and local} NLS equation depending just on
  neighboring lattice points.
At difference from the Ablowitz and Ladik  \cite{al} discrete NLS, the nonlinear term
in equation (\ref{yyyf}) is completely local.
The P$\Delta$E (\ref{yyyf}) has a natural continuous limit when $m_2 \rightarrow \infty$ and  $n_2 \rightarrow \infty$ which, as the coefficients (\ref{yyyy2}) and $\hat c_3$ are real, is just the well known integrable NLS equation.


\section*{Acknowledgments}
The author D.L. thanks the Department of Mathematics,  University of Science and Technology of
China, Hefei (China) for its kind hospitality and the ICTP (Trieste, Italy) which made its visit to China possible by providing a Visiting Scholars/Consultants fellowship.  D.L.  was partially supported by  PRIN Project ``SINTESI-2004'' of the  Italian Minister for  Education and Scientific Research and from  the Projects {\sl Sistemi dinamici nonlineari discreti:
simmetrie ed integrabilit\'a} and {\it Simmetria e riduzione di equazioni differenziali di interesse fisico-matematico} of GNFM--INdAM.


\begin{thebibliography}{0}

\bibitem{al}
M.J. Ablowitz and J.K. Ladik,
{\it Stud. Appl. Math.} {\bf 55}, 213 (1976).

\bibitem{AS} M. Abramowitz and I.A. Stegun,
{\it Handbook of mathematical functions with formulas, graphs, and mathematical tables}
(Dover Publications, Inc., New York, 1992).

\bibitem{calogero}
F. Calogero and W. Eckhaus, {\it Inverse Problems} {\bf 3}, L27 (1987).

\bibitem{Jordan}
C. Jordan, {\it{Calculus of finite differences}} (R\"ottig and Romwalter, Sopron, 1939).

\bibitem{leon}
J. Leon and M. Manna, {\it Jour. Phys. A: Math. Gen.} {\bf 32}, 2845 (1999).

\bibitem{levi}
D. Levi,
{\it Jour. Phys. A: Math. Gen.} {\bf 38}, 7677 (2005).

\bibitem{lh}
D. Levi and H. Heredero,
{\it Jour. Nonlinear Math. Phys.} {\bf 12}, suppl. 1, 440 (2005).

\bibitem{lp}
D. Levi and M. Petrera, math-ph/0510084, submitted to
{\it Jour. Math. Phys.}.

\bibitem{frank1}
R. Sahadevan and H.W. Capel,
{\it Phisica A} {\bf 330}, 373 (2003).

\bibitem{t1}
T. Taniuti,
{\it Prog. Theor. Phys.} {\bf 55 }, 1 (1974).

\bibitem{t2}
T. Taniuti and K. Nishihara,
{\it {Nonlinear waves}} (Monographs and Studies in Mathematics, Pitman, Boston,1983).

\bibitem{ly} 
R. Yamilov, review article in preparation.

\bibitem{kz}
V.E. Zhakarov and E.A. Kuznetsov,
{\it Physica} {\bf 18D}, 455 (1986).


\end{thebibliography}
\end{document}